# Magnetic Structure of $YBaCo_4O_7$ with Kagome- and Triangular-Lattices

Minoru Soda[1], Yukio Yasui[1], Taketo Moyoshi[1], Masatoshi Sato[1], Naoki Igawa[2] and Kazuhisa Kakurai[2]
[1]*Department of Physics, Division of Material Science, Nagoya University, Furo-cho, Chikusa-ku, Nagoya 464-8602*
[2] *Quantum Beam Science Directorate, Japan Atomic Energy Agency, Tokai Ibaraki 319-1195*

Neutron diffraction studies have been carried out in the temperature($T$) range between 10 K and 130 K on a single crystal of $YBaCo_4O_7$, which has the stacking of kagome and triangular lattices of $CoO_4$ tetrahedra along the $c$-axis. Structural transitions have been found at two temperatures $T_{c1}$~70 K and $T_{c2}$~105 K. With decreasing $T$, magnetic order appears along with the transition at $T_{c2}$, but it does not grow to an ideal long-range order even at the lowest $T$ studied (~10 K). Two groups of magnetic diffuse reflections, which originate from the Co-moments of the kagome and triangular lattices, are observed separately in the reciprocal space. With further decreasing $T$, the growth rates of the intensities of these two are enhanced by another transition at $T_{c1}$. These magnetic behaviors seem to be related to the release of the geometrical frustration. At 10 K, the patterns of the magnetic short-range order have been determined for both sites of the kagome and triangular lattices.

corresponding author: e43247a@nucc.cc.nagoya-u.ac.jp

## 1. Introduction

The cobalt oxides $RBaCo_4O_7$(R=Ca, Y and rare earth elements) has a structure with the alternating stack of the kagome and triangular lattices formed of $CoO_4$ tetrahedra[1, 2] (see Fig. 1). Then, a question arises immediately what magnetic behaviors appear in this geometrically frustrated systems. There is another interest related to the spin state of Co ions.

In Co oxides, the spin state often becomes a key element to understand their physical properties. Because the energy differences $\delta E$ among different spin states are small, a wide variety of physical behaviors related to the spin state change may be realized by controlling the $\delta E$ value. To investigate how the spin state depends on their material parameters, authors' group has carried out the neutron diffraction measurements on single crystals of the oxygen deficient perovskite $RBaCo_2O_{5+\delta}$(R=Nd and Tb), having the linkages of $CoO_6$ octahedra and/or $CoO_5$ pyramids, and has also studied transport and magnetic properties on the perovskite $Pr_{1-x}A_xCoO_3$ (A=Ba, Sr and Ca).[3-7] Results of these studies indicate that $\delta E$ or Co spin states can be controlled by changing the ionic radii of $R^{3+}$ and $A^{2+}$ and the average electron number. Not only detailed differences among local structures around Co ions but also the large difference between the pyramidal- and octahedral-arrangements of oxygen atoms affect the $\delta E$ value or the Co spin state through the change of the crystal field strength and/or the electron transfer energy between the Co ions.

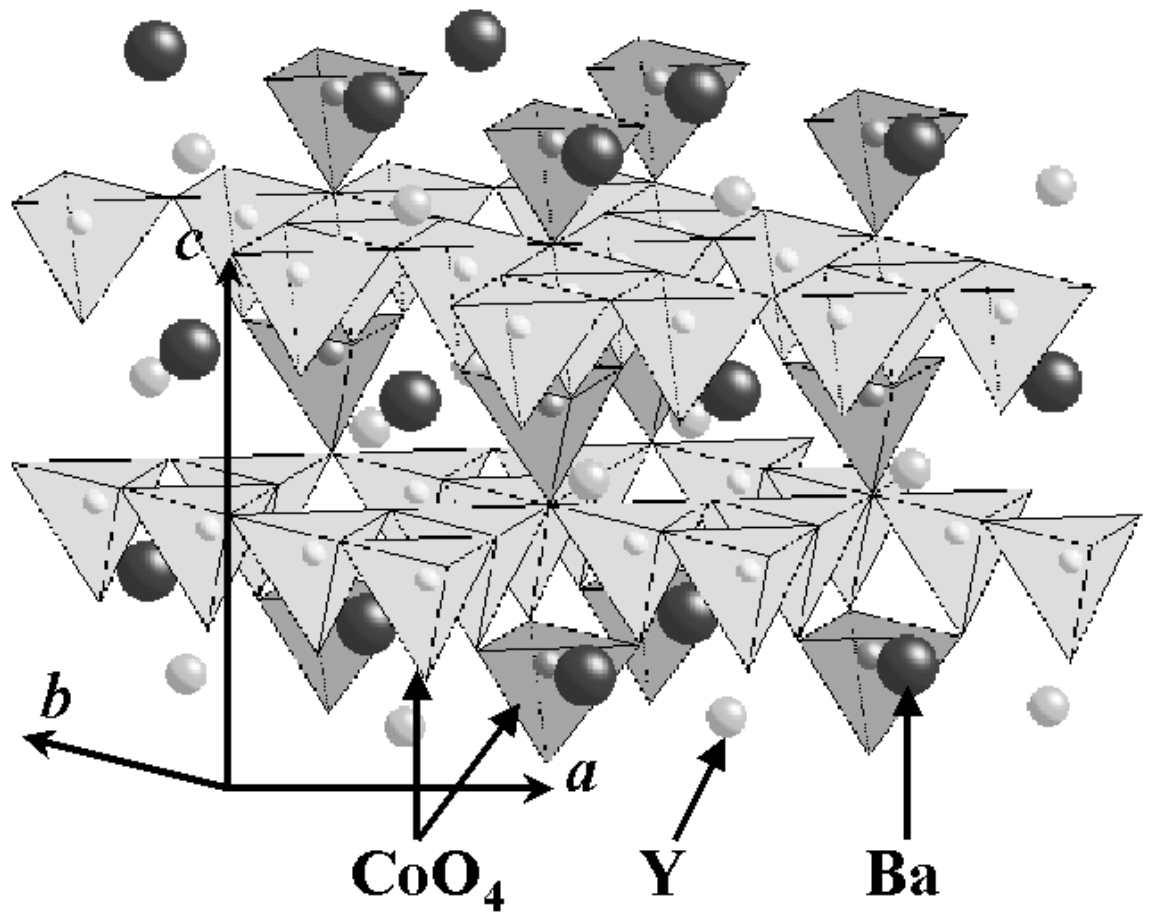

Fig. 1. Schematic structure of $YBaCo_4O_7$.

Because the present system $RBaCo_4O_7$ has the linkages of $CoO_4$ tetrahedra, we can study the spin state of $CoO_4$ tetrahedra and compare it with those of $CoO_5$ pyramid and $CoO_6$ octahedra. It is also interesting to study how the magnetic correlation grows in $RBaCo_4O_7$ with decreasing temperature $T$, because it is expected to have the geometrical frustration.

In the present work, neutron scattering studies have mainly been carried out on a single crystal of $YBaCo_4O_7$ to understand the magnetic correlation at low temperatures. The detailed crystal structure of the system is as follows. It has been reported to be hexagonal (space group; $P6_3mc$), and the lattice parameters are $a=b\sim$ 6.2982 Å and $c\sim$ 10.2467 Å.[1] There are three sorts of equilateral triangles, one in the triangular lattice and two in the kagome lattice with different sizes. The ratio of the Co ion numbers in the kagome- and triangular-lattices is 3 to 1. The ratio of $Co^{2+}$- to $Co^{3+}$-numbers is formally 3 to 1, though it is uncertain that there are charge disproportionations of the Co ions in the kagome and the triangular lattices. The $T$-dependence of the magnetization $M$ suggests that a certain magnetic transition takes place at $T_{c1}$~70 K as described in detail later. Another transition has been found at $T_{c2}$~105 K, too, by measuring the specific heat $C$ and by the neutron diffraction studies.

We also report results of the magnetic structure analyses at 10 K, where those of the kagome- and triangular-lattices have been determined.

## 2. Experiments

Polycrystalline samples of $YBaCo_4O_7$ were prepared by the solid reaction: $Y_2O_3$, $BaCO_3$ and $Co_3O_4$ were mixed with the proper molar ratios and the mixtures were sintered at 1000 °C for 15 h, reground and sintered again at 1150 °C for 20 h. A single crystal of $YBaCo_4O_7$ was grown by a floating zone (FZ) method: The initial mixtures were pressed into rod and sintered at 1000 °C for 15 h. By using the obtained rod, the single crystal was grown. The crystal was annealed at 1000 °C for 60 h in air. Both the polycrystal and the single crystal were checked not to have appreciable amount of impurity phases by powder X-ray measurements. (From the magnetization measurement of a part of the single crystal, the crystal was found to contain about 1~2 % molar fraction of $YBaCo_2O_{5+x}$, which is ferromagnetic in the $T$ region between 150 K and 300 K. [8]) By the thermo gravimetric analyses (TGA) carried out for both the polycrystalline and single crystal specimens, their oxygen numbers were estimated to be 7.15±0.10, and in the present structure analyses, we have assumed that it is 7.0.

The magnetizations $M$ were measured by using a Quantum Design SQUID magnetometer under the magnetic field $H$=1 T in the temperature range of 10-350 K. The magnetic susceptibilities $\chi$(≡$M/H$) taken at $H$=1 T by using a polycrystalline sample and an edge part of the single crystal under the conditions of the zero-field-cooling(ZFC) and field-cooling(FC) are shown against $T$ in Fig. 2. The magnetic transition at $T_{c1}$~70 K can be seen for $H$//$ab$-plane, indicating that the Co-moments are within the $ab$-plane.

The specific heat $C$ was measured by the thermal relaxation method for a polycrystalline sample by using a Quantum Design PPMS. The results are shown in Fig. 3. We can find small anomalies at around two temperatures $T_{c1}$ (~70 K) and $T_{c2}$ (~105 K), where structural transitions were found in the neutron measurements described later.

Neutron measurements were carried out for the single crystal by using the triple axis spectrometer TAS-1 installed at JRR-3 of JAEA in Tokai. The crystal was oriented with the [001] and [100] axes, in one case, and the [001] and [110] ones, in another case, in the scattering plane. The 002 reflections of Pyrolytic graphite (PG) were used for both the monochromator and the analyzer. The horizontal collimations were 40'-40'-80'-open and the neutron wavelength was ~2.359Å Two PG filters were placed in front of the second collimator and after the sample to eliminate the higher order contamination. The sample was set in an Al-can filled with exchange He gas, which was attached to the cold head of the Displex type refrigerator. The ω-scan profiles were used to estimate the integrated intensity of the reflections. The absorption correction of the intensity was made. In the structure analyses, we always used the integrated intensities for both the nuclear and magnetic reflections.

Powder neutron diffraction measurements were carried out by using the high resolution diffractometer HRPD installed at JRR-3. The Ge 331 reflections were used for the monochromator. The horizontal collimations were open(35')-20'-6' and the neutron wavelength was ~1.8233 Å. The sample was packed in a vanadium cylinder (~10 ϕ) and the cylinder was set in an Al-can filled with exchange He gas, which was attached to the cold head of the Displex type refrigerator. Rietveld analyses were carried out by using the Rietan 2000.[9]

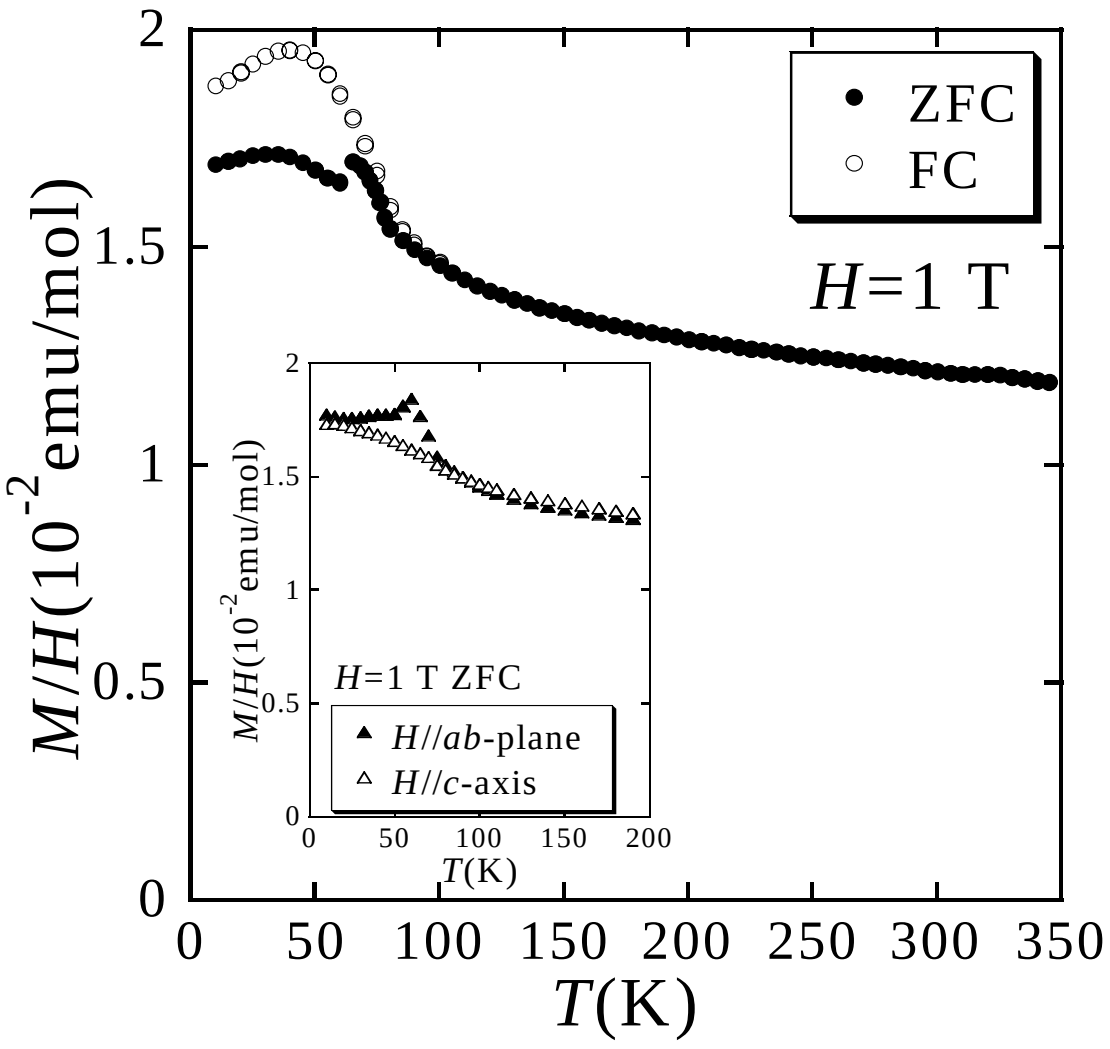

Fig. 2. Temperature dependence of the magnetic susceptibility (≡$M/H$) taken at $H$=1 T with the conditions of zero field cooling (ZFC) and the field cooling (FC). Inset shows the magnetization of a single crystal of $YBaCo_4O_7$ measured with the magnetic field **$H$**//$c$ and **$H$**//$ab$-plane under the condition of ZFC.

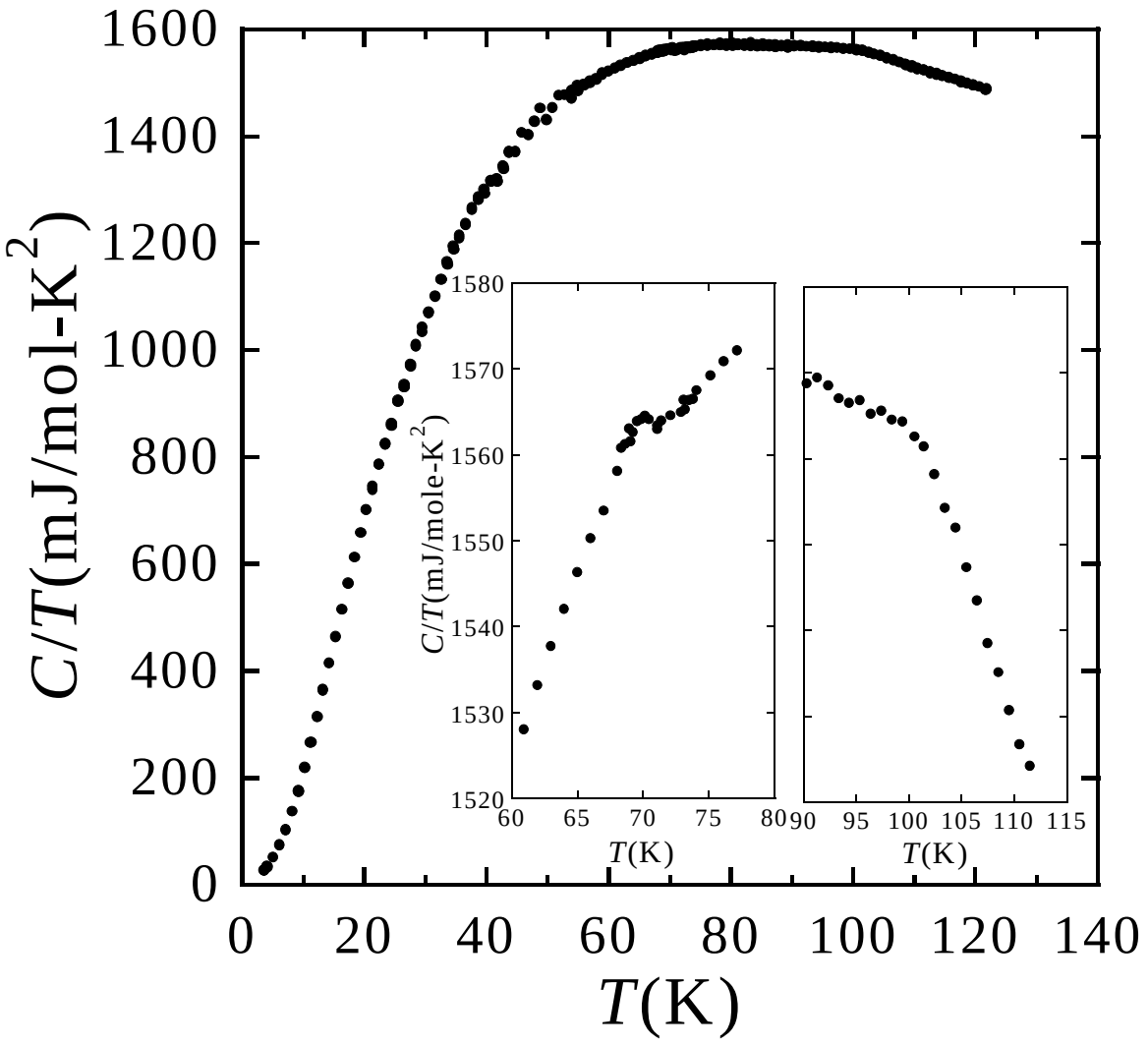

Fig. 3. Temperature dependence of the specific heat divided by $T$ for a polycrystalline sample. Insets show the data around $T_{c1}$ and $T_{c2}$ with the enlarged scales.

### 3. Experimental Results and Discussion

In the neutron studies on the single crystal of $YBaCo_4O_7$, superlattice reflections have been found, besides the fundamental ones, at the ***Q*** points ($h'$/2,0,$l'$) and ($h'$/2,$h'$/2,$l'$) ($h'$=odd; hereafter, we call these points ***Q****-group A*) and ($h''$/3,$h''$/3,$l''$) ($h''$/3=non-integer number; hereafter we call these points ***Q****-group B*) in the reciprocal space of the hexagonal unit cell (*a*=*b*~6.2982 Å, *c*~10.2467 Å). Examples of observed ω-scan profiles of these reflections are shown in Fig. 4 at various *T* values. We have found that among the reflections at the ***Q****-group B*, only the Gaussian component is observed for odd *l*" values and that other superlattice reflections seem to consist of two components, Gaussian and Lorentzian ones. Then, assuming that they consist of a (narrow) Gaussian component and a (broad) Lorentzian one, we have carried out the profile-fittings, and found following things.

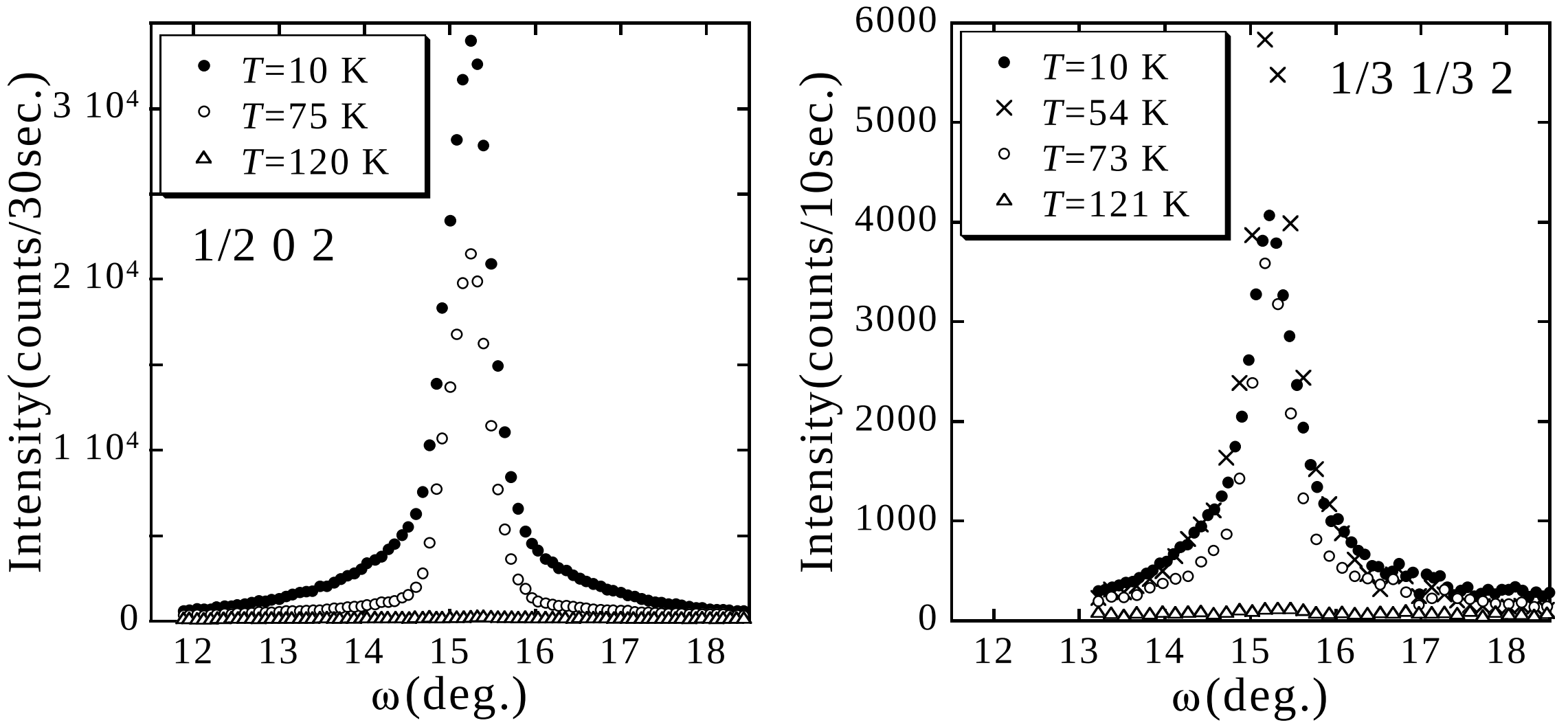

Fig. 4. Examples of the ω-scan profiles of the superlattice reflections obtained at various values of *T*. Inset shows the fundamental reflection, whose |***Q***| value is close to the superlattice reflections.

From the |***Q***|-dependence of the integrated intensities, the Gaussian- and Lorentzian-components can be considered to have the nuclear and magnetic origins, respectively. (The fundamental reflections have only Gaussian component.) As stated above, there exists only the nuclear contribution at the odd *l*" points of ***Q****-group B* (***Q****-group* $B_1$) and of course at the fundamental Bragg points. At the ***Q****-group A* and at the even *l*" points of ***Q****-group B* (***Q****-group* $B_2$), there are both nuclear and magnetic reflections. Detailed magnetic structure analyses described later show that the Co-moments on the kagome- and triangular-lattices can be considered to order with the periods corresponding to the ***Q****-group A* and $B_2$, respectively.

The integrated intensities of the Gaussian (nuclear) and Lorentzian (magnetic) components of the superlattice reflections determined by the fittings are shown in Fig. 5 against *T*, where the total intensities of the reflections are also plotted. The Gaussian components at the ***Q****-group A* appear at $T_{c2}$~105 K with decreasing *T*, and their widths are resolution limited, while at the ***Q****-group B*, the Gaussian components appear at $T_{c1}$~70 K with decreasing *T*, their widths are also resolution limited. As mentioned above, the temperatures $T_{c1}$ and $T_{c2}$ are in rough agreement with those determined by the specific heat measurements. We think it natural to expect that the appearance of the Gaussian components at the ***Q****-group A* are related with the distortion of the kagome lattice and those at the ***Q**-group B* originate from the distortion of the triangular lattice. (We can confirm these presumption by the fact that the Gaussian superlattice components at the ***Q**-group A* between $T_{c1}$ and $T_{c2}$ can be well explained by a space group with only the distortion of the kagome lattice, as described later. The rapid growth of the magnetic correlation observed at around $T_{c1}$ with decreasing $T$ at the ***Q**-group $B_1$* also suggests that the distortion at $T_{c1}$ is mainly related with the triangular lattice, because the distortion of the triangular lattice is expected to primarily affect the correlation of the Co moments on the triangular lattice.)

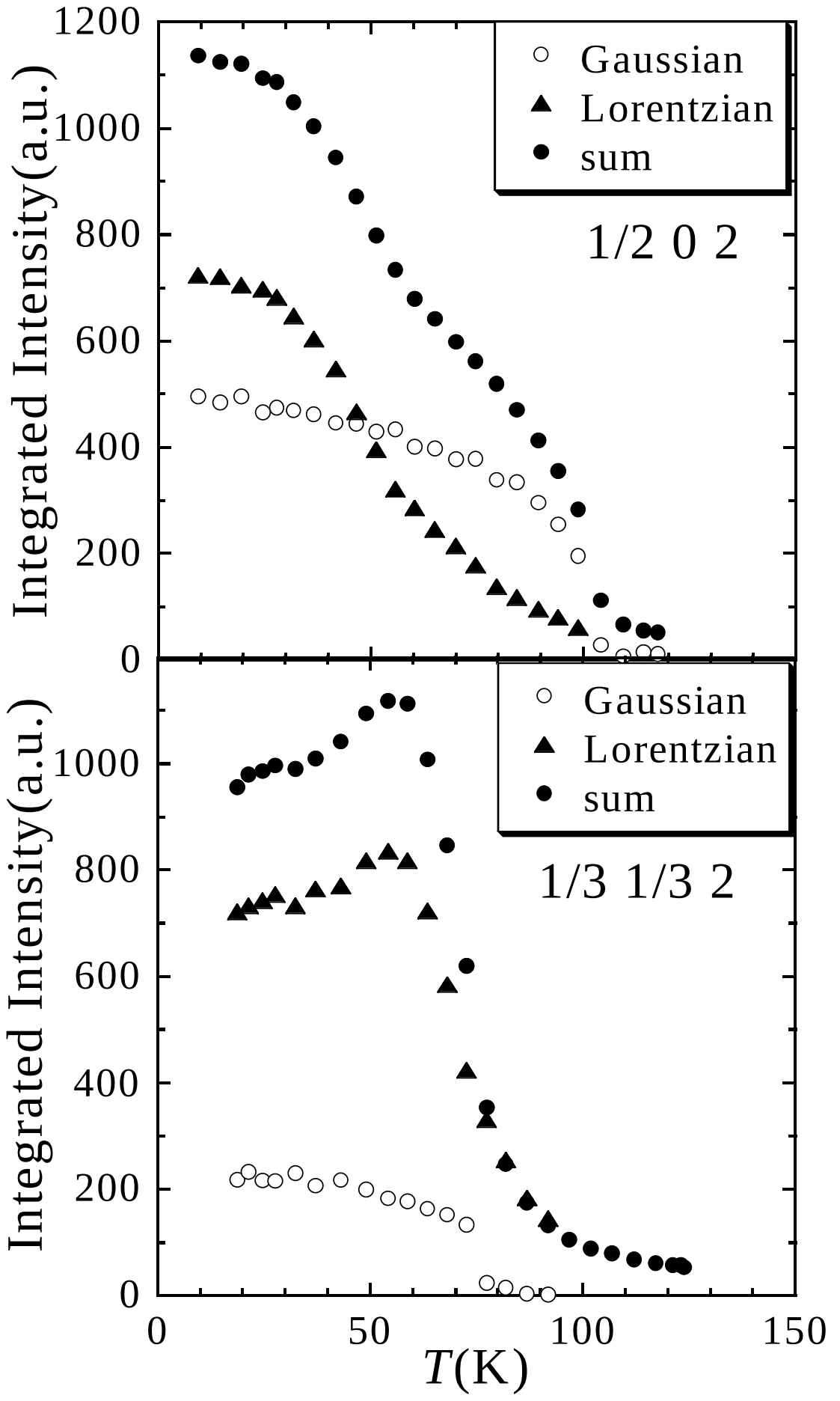

Fig. 5. *T* dependences of the integrated intensities of Gaussian components (nuclear reflections) and Lorentzian ones (magnetic reflections) obtained by fitting the calculated curves to the ω-scan profiles. The total integrated intensity of the superlattice reflections are also plotted by closed circles.

The Lorentzian (magnetic) components due to the ordering (or in the strict sense, due to the growth of the short range correlation) of the Co moments on the kagome lattice appear gradually below $T_{c2}$. The superlattice reflections due to the ordering (or in the strict sense, due to the growth of the short-range correlation) of Co moments on the triangular lattice begin to grow gradually at around $T_{c2}$, too. The short-range magnetic correlation does not reach the long-range one even at $T_{c1}$. Even at 10 K, the correlation length is finite (~150 Å).

With decreasing $T$, the structural change at $T_{c2}$, more or less, releases the geometrical frustration in the kagome lattice, and the magnetic correlation grows gradually below this temperature. In the triangular lattice, the magnetic correlation is gradually induced by the magnetic short-range order in the kagome lattice below $T_{c2}$. The structural transition at $T_{c1}$ enhances the growth rate of the magnetic ordering in both the kagome- and triangular-lattices.

It should be noted, here, that 001, 003, and a few other nuclear reflections (Gaussian components) with indices $h'/2h'/2l'$ ($h'$=odd, $l'$=odd) and those at the ***Q**-group $B_1$*, have been observed even above 105 K, though they are forbidden for the reported space group of $P6_3mc$,[1, 2] We have neglected these reflections in the crystal structure analyses described below, because they are weaker than the other ones (the strongest is just less than ~ 1/10 of the strongest superlattice reflection below $T_{c2}$). At this moment, we do not distinguish if these reflections exist only in the present single crystal or they exist in sintered samples, too. (In refs. 1 and 2, a powder sample was used. Single crystals prepared from the molten phase may have a slightly different structure from that of sintered samples, because lattice imperfections such as inhomogeneity may be introduced in the course of the crystal growth.) The excess oxygens may be another origin of the reflections forbidden for $P6_3mc$.

Before analyzing the magnetic structures of the present system, the crystal structure was studied by using the data on the present single crystal. The integrated intensities of the fundamental reflections at 120 K have been analyzed by adopting the space group $P6_3mc$,[1, 2] where the absorption correction was made, and as stated above, the 001, 003, $h'/2h'/2l'$ ($h'$=odd, $l'$=odd) reflections and those at the ***Q**-group $B_1$* have been omitted. The lattice parameters are $a$=$b$~ 6.2973 Å and $c$~ 10.2218 Å. Results of the fitting of the calculated intensities ($I_{cal}$) of the nuclear reflections to the observed intensities ($I_{obs}$) are shown in Fig. 6, where the values of $I_{obs}$ are plotted against $I_{cal}$. The positional parameters obtained by the fitting are basically consistent with those ever reported.[1)]

In the region $T_{c1} < T < T_{c2}$, the observed set of the superlattice reflections indicates that the proper space group is $Pmc2$ (orthorhombic), where the unit cell parameters are $a \cong a_h$, $b \cong \sqrt{3} \times a_h$ and $c \cong c_h$, $a_h$ and $c_h$ being the parameters of the hexagonal cell. For this space group, the triangles of both kinds of lattice do not necessarily be equilateral. We have not carried out detailed structure analyses on the single crystal below $T_{c1}$.

Powder neuron diffraction measurements were also carried out on polycrystalline sample. Because the superlattice reflections are very week, as suggested by the single crystal studies, they could not be observed even in the distorted phases at 10 K and 85 K except 1/202 (or 1/31/32) reflections. At 120 K, we have observed no superlattice peaks, and the crystal structure has been found to be well explained by the space group P6$_3$*mc*. At *T*~85 K, the Rietveld analyses by using the space groups P*mc*2 (orthorhombic) and P6$_3$*mc* (hexagonal) have been carried out. The result for the former space group is shown in Fig. 7. (The region of small 2θ was omitted in the analysis to ensure that magnetic reflections are not included.) The $R_{WP}$-values for P*mc*2 and P6$_3$*mc* are 6.94 % and 8.43 %, respectively. It probably indicates that the system has the space group P*mc*2 in the *T*-region $T_{c1} < T < T_{c2}$. The structure could not be clarified at 10 K and microscopic details on the structural change at $T_{c1}$, which mainly associated with the triangular lattice, remains unsolved.

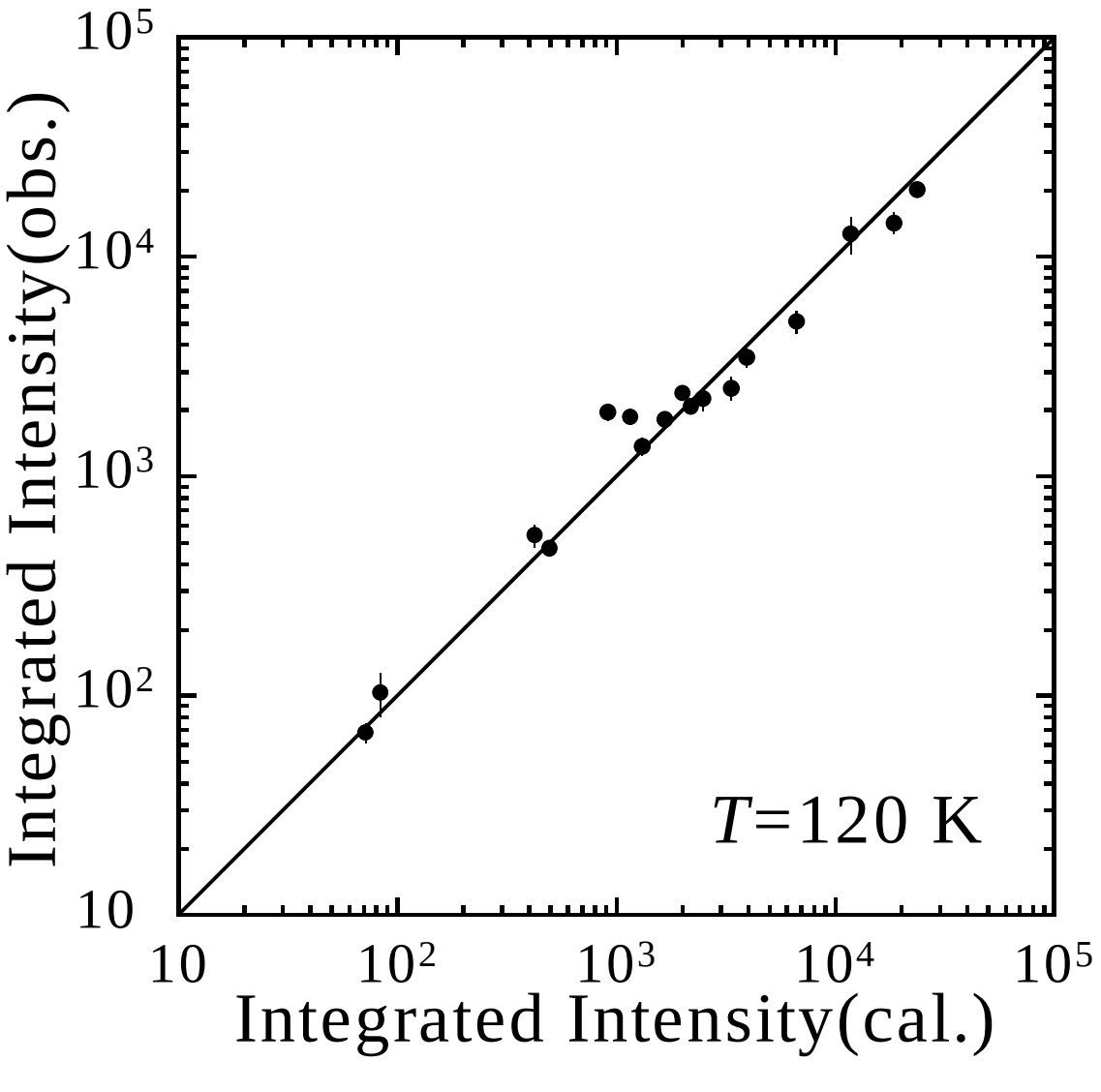

Fig. 6. Integrated intensities of the nuclear Bragg reflections collected at 120 K are plotted against the calculated values. The space group $P6_3mc$ is used in the calculation.

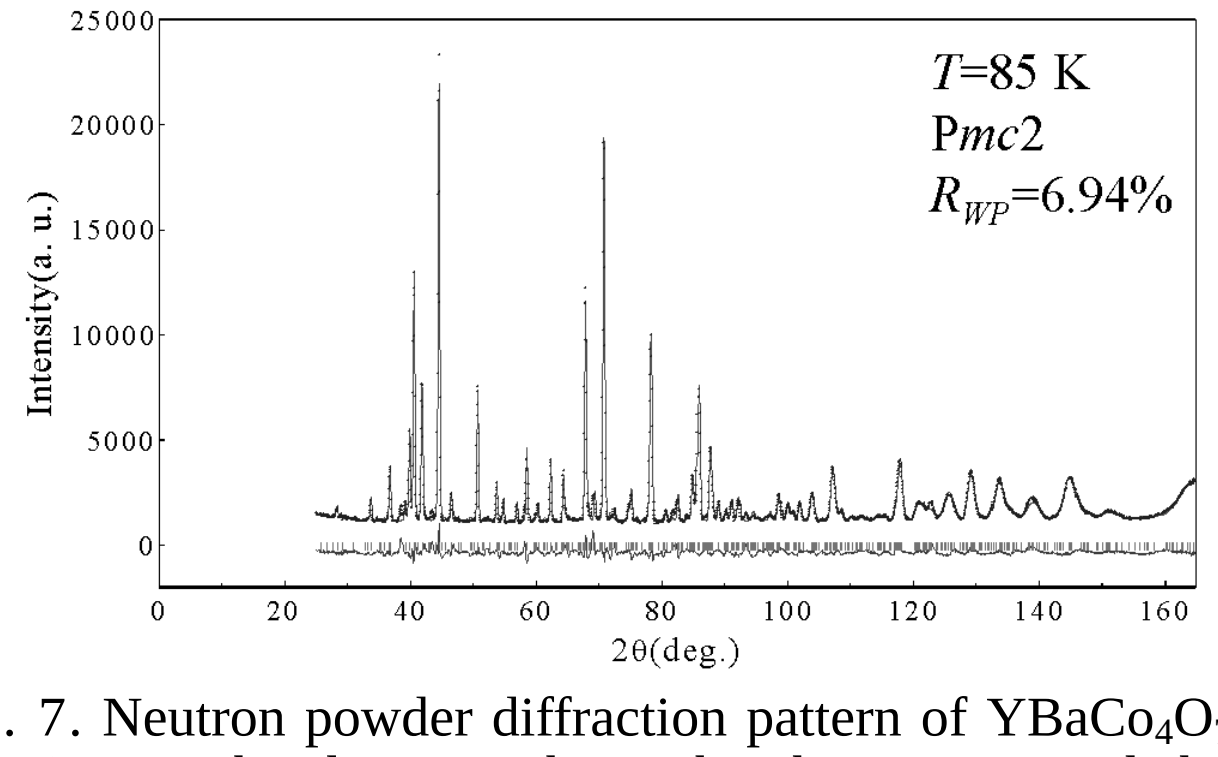

Fig. 7. Neutron powder diffraction pattern of $YBaCo_4O_7$ at 85 K. The data are shown by the crosses and the calculated curve obtained by the Rietveld fitting with orthorhombic ($Pmc2$) structure is shown by the solid curve. The Bragg positions are indicated by the vertical lines and the difference between the observed and calculated data are also shown by the gray line.

In the magnetic structure analyses of the present single crystal, the Co positions at 120 K (space group P6$_3$*mc*) have been used. It is rationalized for the following reasons. The magnetic scattering components (Lorentzian one) can be separated from the nuclear ones (Gaussian ones) and treated almost independently, because the small distortions do not bring about significant effects on their intensities.

The intensities of the Lorentzian components have been used as those of the magnetic scattering. The magnitudes of the magnetic moments at crystallographically equivalent sites are assumed to be equal. For the magnetic form factors, the averaged values of the isotropic ones of $Co^{3+}$ and $Co^{2+}$ reported in ref. 10 were used. In the fittings, the volume ratios of (*h*,*h*,*l*), (-*h*,2*h*,*l*) and (2*h*,-*h*,*l*) domains or (*h*,0,*l*), (0,*h*,*l*) and (-*h*,*h*,*l*) domains are assumed to be 1:1:1.

In the detailed calculations, we have treated, as is already stated in the analyses of the Gaussian components,

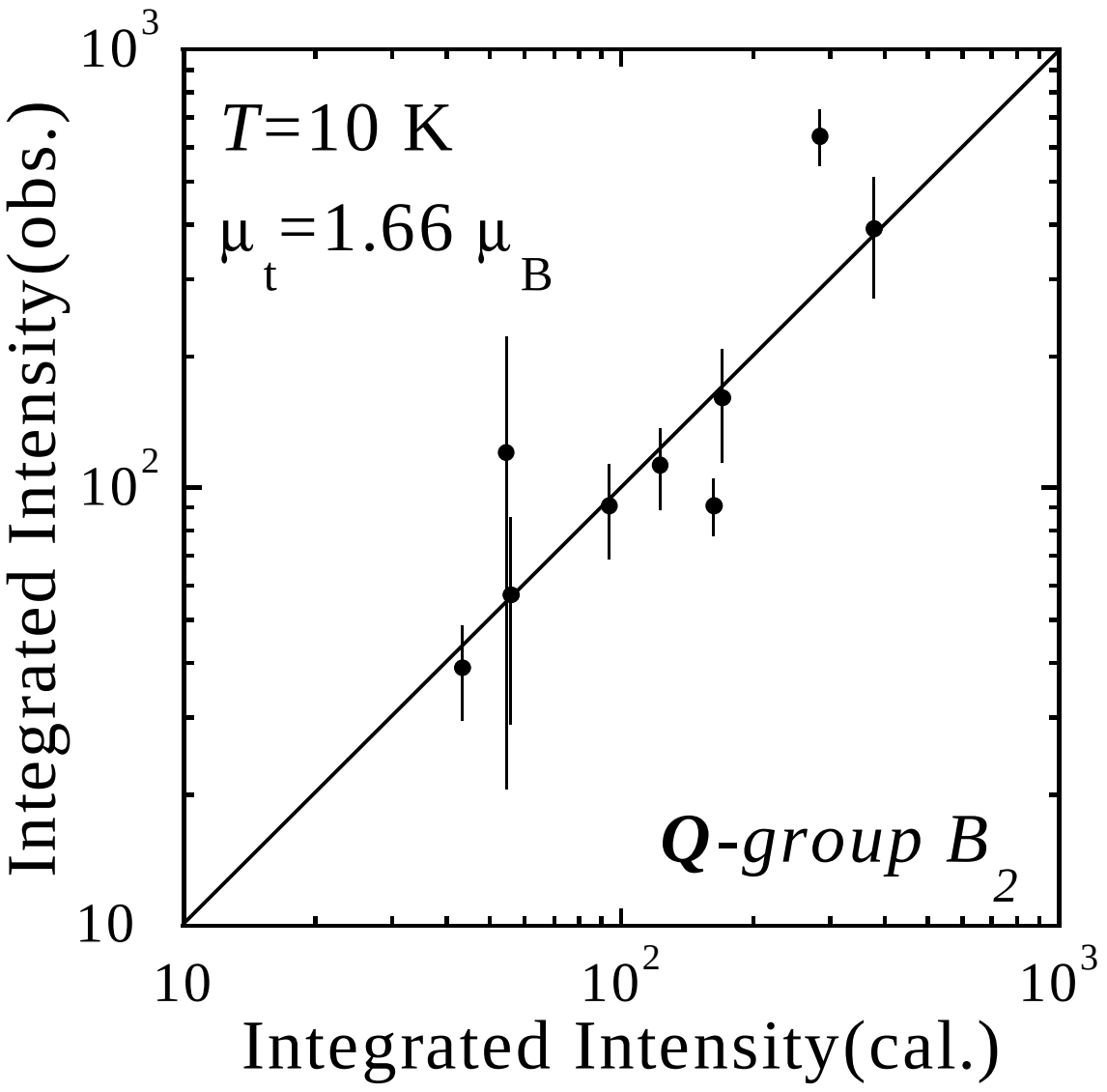

Fig. 8. Integrated intensities of the magnetic reflections from the moments on the triangular lattice collected at 10 K are shown against those of the model calculation for the 120° structure.

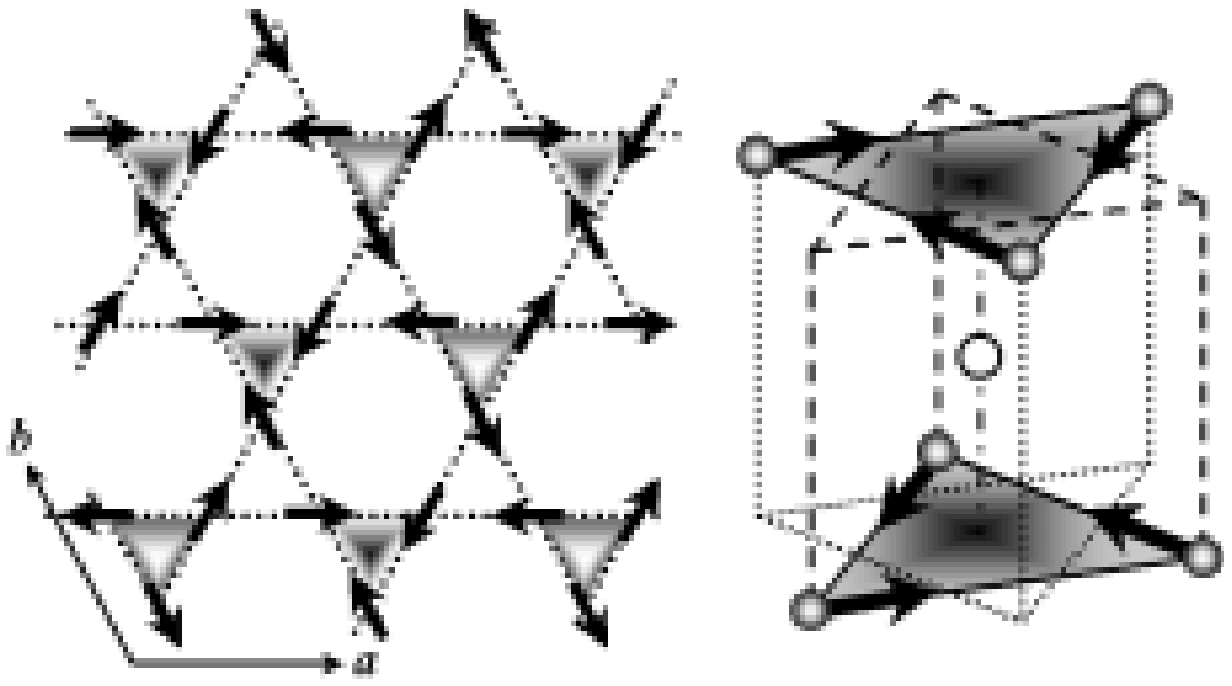

Fig. 9. Left: Magnetic structure of the kagome lattice of $YBaCo_4O_7$ at 10 K. Right: Geometrical relationship between the triangular- and adjacent kagome-lattices. The open circle indicates a Co ion of the triangular lattice between the kagome lattices formed of the Co ions shown by the gray circles.

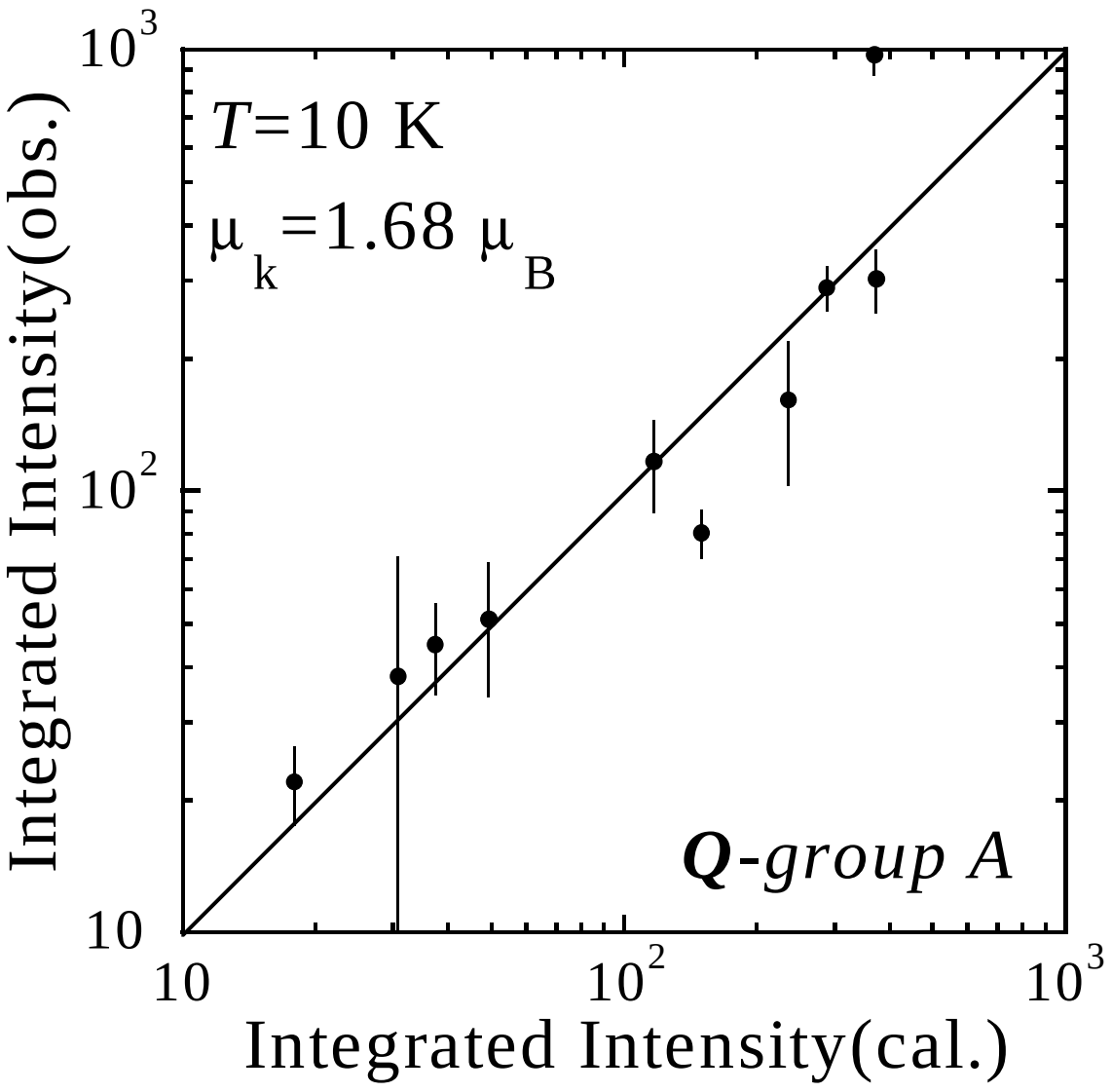

Fig. 10. Integrated intensities of the magnetic reflections of the kagome lattice collected at 10 K are plotted against those calculated by using the magnetic structure shown in Fig. 9.

the two groups of the Lorentzian superlattice components at the ***Q****-group A* and $B_2$, independently, as the contributions of the moments on the kagome- and triangular-lattices, respectively, because other possibilities are very difficult to adopt. For example, if the Co moments on the kagome lattice contribute to the reflections at the ***Q****-group B*, the existence of Lorenzian components are expected to appear at the ***Q****-group* $B_1$, but they do not.

The Co-moments on the triangular lattice have the so-called 120° structure, and the moments at the sites shifted ±*c*/2 along the *c* direction are parallel to those of the original sites. For the 120° structure, the direction of the Co-moments cannot be determined. (We can only say that they are within the plane.) The integrated intensities of the magnetic reflections of the triangular lattice collected at 10 K are shown in Fig. 8 against the calculated intensities. The magnitude $\mu_t$ of the Co-moments on the triangular lattice is 1.66(±0.11) $\mu_B$.

The magnetic structure of the kagome lattice obtained at 10 K is shown schematically in the left part of Fig. 9. Spins on the triangular lattice are located at the positions shifted by ~±*c*/4 along the *c* direction from the center of gravity of the shaded triangles of the kagome lattice, as shown in the right part of Fig. 9. The integrated intensities of the magnetic reflections of the kagome lattice collected at 10 K are shown in Fig. 10 against the values calculated for the magnetic structure drawn in Fig. 9. The magnitude $\mu_k$ of the Co-moments on the kagome lattice is 1.68(±0.15) $\mu_B$. Here, we realize that the relation $\mu_k \cong \mu_t$ holds, though the magnetic structures of the kagome- and the triangular-lattices have been analyzed separately. It indicates that there is no charge disproportionation of Co ions. Although the Co ions may be considered, from the magnitudes of Co-moments, to be in the low spin- or intermediate spin-state, we cannot determine the states unambiguously, because the spin system is not completely ordered even the lowest temperature studied here (~ 10 K).

Now, we have succeeded in the magnetic structure analyses of $YBaCo_4O_7$, by treating the two groups of the magnetic superlattice reflections, independently. The Co-moments on the triangular lattice have the 120° structure, while the 120° structure is found only for the shaded triangles in the kagome lattice. It may indicate that the release of the geometrical frustration occurs by the structural change.

## 4. Conclusion

Neutron diffraction studies have been carried out on a single crystal of $YBaCo_4O_7$, which has the linkages of $CoO_4$ tetrahedra, in the temperature range between 10 K and 130 K. Two structural phase transitions have been found at $T_{c1}$~70 K and $T_{c2}$~105 K. The Co moments are also involved in these transitions, probably through the release of the geometrical frustration. Although they do not exhibit the long-range order down to the lowest temperature studied here (~10 K), the strong diffuse contributions due to the short-range order of the moments on the kagome and triangular lattices have been observed. Because the ***Q*** regions of the scatterings from these two lattices are different, we can analyze their intensities separately. We have found that the moments on the triangular lattice have the 120° structure and that the structure of the kagome lattice is closely related with the 120 ° structure.

Acknowledgments - Work at the JRR-3 was performed within the frame of JAEA Collaborative Research Program on Neutron Scattering. The work is supported by Grants-in-Aid for Scientific Research from the Japan Society for the Promotion of Science (JSPS) and by Grants-in-Aid on priority area from the Ministry of Education, Culture, Sports, Science and Technology.

1) M. Valldor and M. Andersson: Solid State Sci. **4** (2002) 923.
2) M. Valldor: Solid State Sci. **6** (2004) 251.
3) M. Soda, Y. Yasui, T. Fujita, T. Miyashita, M. Sato and K. Kakurai: J. Phys. Soc. Jpn. **72** (2003) 1729.
4) M. Soda, Y. Yasui, M. Ito, S. Iikubo, M. Sato and K. Kakurai: J. Phys. Soc. Jpn. **73** (2004) 464.
5) M. Soda, Y. Yasui, M. Ito, S. Iikubo, M. Sato and K. Kakurai: J. Phys. Soc. Jpn. **73** (2004) 2857.
6) T. Fujita, T. Miyashita, Y. Yasui, Y. Kobayashi, M. Sato, E. Nishibori, M. Sakata, Y. Shimojo, N. Igawa, Y. Ishii and K. Kakurai, T. Adachi, Y. Ohishi and M. Takata: J. Phys. Soc. Jpn. **73** (2004) 1987.
7) T. Fujita, S. Kawabata, M. Sato, N. Kurita, M. Endo and Y. Uwatoko: J. Phys. Soc. Jpn. **74** (2005) 2294.
8) D. Akahoshi and Y. Ueda: J. Solid State Chem. **156** (2001) 355.
9) F. Izumi and T. Ikeda: Mater. Soc. Forum **321-324** (2000) 198.
10) P. J. Brown: *International Tables for Crystallography*, ed. by A. J. C. Wilson (Kluwer, Dordrecht, 1992) vol. C, chap. 4